\documentclass[preprint,aps,showpacs,nofootinbib,preprintnumbers,amsmath,amssymb]{revtex4}
\usepackage{amssymb}
\usepackage{}
\usepackage{epsfig}
\usepackage{graphicx}% Include figure files
\usepackage{subfigure}
\usepackage{dcolumn}% Align table columns on decimal point
\usepackage{bm}% bold math
\usepackage[usenames ,dvipsnames]{xcolor}
\usepackage{slashed}

\begin{document}

\title{Study of  Two-Loop Neutrino Mass Generation Models}

\author{Chao-Qiang~Geng$^{1,2,3}$\footnote{geng@phys.nthu.edu.tw} and
Lu-Hsing Tsai$^{2}$\footnote{lhtsai@phys.nthu.edu.tw}}
  \affiliation{
 $^{1}$Chongqing University of Posts \& Telecommunications, Chongqing, 400065, China\\
 $^{2}$Department of Physics, National Tsing Hua University, Hsinchu, Taiwan\\
  $^{3}$Physics Division, National Center for Theoretical Sciences, Hsinchu, Taiwan}

\date{\today}
\begin{abstract}
We study  the models with the Majorana neutrino masses  generated radiatively by two-loop diagrams due to the Yukawa 
 $\rho \bar \ell_R^c \ell_R$  and  effective  $\rho^{\pm\pm} W^\mp W^\mp$ couplings along with a scalar triplet $\Delta$,
where $\rho$ is a doubly charged singlet scalar, $\ell_R$ the charged lepton and $W$ the charged gauge boson.
A generic feature in these types of models is that the neutrino mass spectrum has to be a normal hierarchy.
Furthermore, by using the neutrino oscillation data and comparing with the global fitting result in the literature,  
we find  a unique neutrino mass matrix  and predict the Dirac and two Majorana CP phases to be $1.40\pi$, 
$1.11\pi$ and $1.47\pi$, respectively. We also discuss the model parameters constrained 
by the lepton flavor violating processes and electroweak oblique parameters. In addition, we show that the rate of the 
neutrinoless double beta decay $(0\nu\beta\beta)$ can be as large as the current experimental bound as it is dominated 
by the short-range contribution at tree level, whereas the traditional long-range one is negligible. 
\end{abstract}

%\pacs{95.35.+d, 98.70.Sa, 13.85.Tp, 14.80.-j}
%\keywords{}
\maketitle
\section{Introduction}
Although the data from neutrino experiments  have implied  that at least two neutrinos carry nonzero  
masses~\cite{Anselmann:1992kc,Fukuda:1998mi,Ahmad:2002jz,Ahmad:2002ka,Ahn:2006zza},  the origin of these masses 
is still a mystery. 
Apart from the mass generation of Dirac neutrinos given by the Yukawa couplings with the existence of right-handed neutrinos ($\nu_R$),
 seesaw mechanisms with type-I~\cite{TypeIseesaw1,TypeIseesaw2,TypeIseesaw3,TypeIseesaw4,TypeIseesaw5}, type-II~\cite{typeIIseesaw1,typeIIseesaw2,typeIIseesaw3,typeIIseesaw4,typeIIseesaw5,typeIIseesaw6,typeIIseesaw7} and type-III~\cite{Foot:1988aq} can generate  masses  for Majorana neutrinos by realizing the Weinberg operator $(\bar L_L^c\Phi)(\Phi^TL_L)$ at tree-level, where
 $\Phi$ and $L_L$ are the doublets of  Higgs and left-handed lepton fields, respectively. In these scenarios, 
either heavy degrees of freedom or tiny coupling constants  are required in order to conceive  the small neutrino masses. 
On the other hand, models with the Majorana neutrino masses generated at one-loop~\cite{Zee:1980ai,Ma:2006km}, two-loop~\cite{Zee:1985id, Babu:1988ki,Chen:2006vn,Chen:2007dc} and higher loop~\cite{Krauss:2002px, Aoki:2008av, Gustafsson:2012vj,Geng:2014gua} diagrams have also been proposed without introducing $\nu_R$.
Due to the loop suppression factors, the strong bounds on the coupling constants 
  and  heavy states are relaxed, resulting in a somewhat natural  explanation for the smallness of neutrino masses.

Among the loop-level mass generation mechanisms, there is a special type of the neutrino models~\cite{Chen:2006vn,Chen:2007dc}
 in which a doubly charged singlet scalar $\rho:(1,4)$
 and a triplet $\Delta:(3,2)$\footnote{The convention for the electroweak quantum numbers $(I,Y)$ with $Q=I+{Y/2}$ is used throughout this paper.} under $ {\rm SU}(2)_L\times {\rm U}(1)_Y$ are introduced 
to yield the new Yukawa coupling $\rho \bar \ell_R^c \ell_R$ with the charged lepton $\ell_R$ as well as 
the effective gauge coupling $\rho^{\pm\pm} W^\mp W^\mp$ due to the mixing between $\rho^{\pm\pm}$ and $\Delta^{\pm\pm}$,
leading to the neutrino masses through two-loop diagrams~\cite{Chen:2006vn}.
As this model is the simplest way to realize the $\rho WW$ coupling, we name it as the minimal two-loop-neutrino model (MTM)~\cite{Chen:2006vn}. 
%, 
 It is interesting to note that $\rho^{\pm\pm} W^\mp W^\mp$ can also be induced from non-renomalizable 
   high-order operators~\cite{Gustafsson:2014vpa,King:2014uha, Sierra:2014rxa}. 
Although MTM can depict neutrino masses at two-loop level, 
the assumption on the absent of the  $\bar L^cL\Delta$ term makes this model unnatural. 
 To solve this problem, one can simply extend MTM by adding an extra doublet scalar, which together with $\Delta$
 carries an odd charge under an $Z_2$ symmetry~\cite{Chen:2010ir}.
We  call this model as the doublet  two-loop-neutrino model~(DTM).
 On the other hand,
 $\rho^{\pm\pm} W^\mp W^\mp$ could  be granted by inner-loop diagrams, such as those~\cite{Gustafsson:2012vj,Geng:2014gua}
 with  three-loop contributions to neutrino masses, in which the neutral particle in the inner-loops could be  a candidate for the stable dark matter.

In this study, we will demonstrate that 
the neutrino mass matrix can be determined 
in these models by the experimental data. In particular, the neutrino mass spectrum is found to be a normal hierarchy. 
In addition, the neutrinoless double beta decay $(0\nu\beta\beta)$ is dominated by the short-range contribution at tree level
due to the effective coupling of $\rho^{\pm\pm} W^\mp W^\mp$~\cite{Chen:2006vn,Chen:2007dc,Gustafsson:2014vpa,Geng:2014gua,King:2014uha,delAguila:2011gr, delAguila:2012nu}, instead of the traditional long-range one. However, the neutrino masses in this type of the models are usually over suppressed as  there is not only a two-loop suppression factor, but also a small ratio
$m_l/v$ with the charged lepton mass $m_l$ and vacuum expectation value (VEV) $v=246$ GeV of the Higgs field.
Furthermore, the lepton flavor violation  (LFV) processes 
 could also limit  the new Yukawa couplings. To have a large enough neutrino mass, the mixing angle or  mass splitting between the two doubly-charged states should be large, which  
 inevitably leads to a significant contribution to the electroweak oblique parameters, especially the $T$ parameter. We will calculate the neutrino masses in details and check whether there is a  tension between these masses and  the constraint from the oblique parameter $T$.

This paper is organized as follows. In Sec II, we study the neutrino masses in 
the two-loop neutrino models. 
In Sec III, the constraints on the model parameters  from  lepton flavor violating processes
and electroweak oblique parameters are studied.  We  present the conclusions in Sec. IV.

\section{Two-loop Neutrino Masses}
In MTM, we introduce  the scalars $\rho:(1,4)$ and $\Delta=(\Delta^{++},\,\Delta^+,\,\Delta^0):(3,2)$
under $SU(2)_L\times U(1)_Y$. 
The relevant terms in the Lagrangian are given by
\begin{eqnarray}
-{\cal L}&=&-\mu_\Phi^2(\Phi^\dagger\Phi)+M_\Delta^2 (\Delta^\dagger\Delta)+\lambda_\Phi(\Phi^\dagger\Phi)^2+\bar \lambda_3(\Delta^\dagger\Delta)_1(\Phi^\dagger\Phi)_1+\bar \lambda_4(\Delta^\dagger\Delta)_3(\Phi^\dagger\Phi)_3\nonumber\\
&&+\left[Y_{ab} {(\bar L_L^c)_{a}}\Delta (L_{L})_b+{C_{ab}\over2}\rho {(\bar \ell_R^c)_{a}} (\ell_{R})_b-\mu \Delta (\Phi^\dagger)^2+{\kappa\over2}\rho^*\Delta^2+ \bar \lambda \rho^* \Delta \Phi^2 +{\rm H.c.}\right]\,,\label{Eq_TripletLagrangian}
\end{eqnarray}
where $\Phi=(\Phi^+,\,\Phi^0)^T$ with $\Phi^0=(\Phi_R+i\Phi_I)/\sqrt{2}$ is the SM doublet scalar, the indices of  $a$ and $b$ 
represent $e$, $\mu$ and $\tau$, and  the subscripts of $1$ and $3$  in the quartic terms 
stand for the SU(2) singlet and triplet scalars inside the parentheses, respectively. 
After the spontaneous symmetry breaking, $\Phi$ acquires a VEV of $v_\Phi=\sqrt{2}\left<\Phi^0\right>$, while 
the neutral component of $\Delta$ also receives a VEV $v_\Delta/\sqrt{2}$,  generated via the $\mu$ term.
Note that by the global fitting result of $\rho_0=1.0000\pm0.0009$~\cite{Agashe:2014kda}, $v_\Delta$ is constrained to be $\lesssim5\,{\rm GeV}$, 
so that $v_\Phi\simeq246\,{\rm GeV}$ is a good approximation. The $\kappa$ term in Eq.~(\ref{Eq_TripletLagrangian})
can produce a mixing term between $\rho^{\pm\pm}$ and $\Delta^{\pm\pm}$, resulting in two mass eigenstates $P_{1,2}$ with masses $M_{1,2}$, respectively. 
We will set $Y_{ab}$ and $\bar \lambda$ to be 
zero since
they have the tree-level and logarithmic divergent two-loop contributions to
neutrino masses, respectively.
These two coupling can also be forbidden in a natural way by introducing a new doublet~\cite{Chen:2010ir} or
 a singlet scalar~\cite{delAguila:2011gr} with an $Z_2$ symmetry or by replacing $\Delta$
by a higher multiplet, such as $\xi:(5,2)$ without the discrete symmetry~\cite{Chen:2012vm}. 
 The scalar mass spectra of MDM are shown in Appendix \ref{App_Mass}.1.

\begin{figure}
\includegraphics[width=12cm]{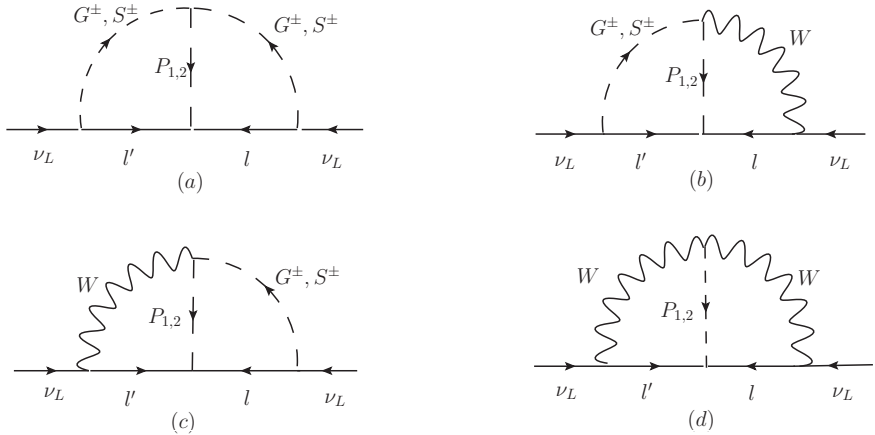} 
\caption{Diagrams for the neutrino mass generation, where the charged states $S^\pm$ can be replaced by $S_1^\pm$ or $S_2^\pm$ when DTM is discussed.}
\label{Fig_NeutrinoMass}
\end{figure}

We now calculate the neutrino masses from the two-loop diagrams of Fig.~\ref{Fig_NeutrinoMass}
in the t'Hooft-Feynman gauge. 
The neutrino mass matrix $M_\nu$ can be written as
\begin{eqnarray}
(M_\nu)_{\ell\ell'}&=&{1\over (16\pi^2)^2}{2C_{\ell\ell'}m_{\ell}m_{\ell'}\over v^2}\left[A_{(a)}+A_{(b)}+A_{(c)}+A_{(d)}\right]\,,
\label{Eq_neutrinomassMTM}
\end{eqnarray}
where the integration results $A_{(a)}$, $A_{(b)}$, $A_{(c)}$ and $A_{(d)}$, 
corresponding to the sub-figures (a), (b), (c), and (d) in Fig.~\ref{Fig_NeutrinoMass},   are given  in Eqs.~(B9)-(B11) in Appendix~\ref{App_neutrinomass}, respectively. 
Explicitly, we find that the contribution related to $A_{(a)}$ dominates over the other three components. We note  
that if $M_\rho$ is much smaller than $M_\Delta$ and the mixing angle $\theta$ between them is small,  
this model approximately coincides with 
the effective theory involving the dimension-7 operator $\rho (D_\mu \Phi)(D^\mu \Phi)\Phi\Phi$ discussed in Ref.~\cite{King:2014uha}. 

DTM can be viewed as the extension of  MTM 
by introducing a new doublet $\chi=(\chi^+,\,\chi^0)^T$ with $\chi^0=(\chi_R+i\chi_I)/\sqrt{2}$.
This new doublet along with  $\Delta$  carries an odd charge under the $Z_2$ symmetry~\cite{Chen:2010ir}. 
This discrete symmetry can forbid the tree-level coupling $\bar L^c L \Delta$ to make the two-loop neutrino mass generation more natural. 
The relevant part of the Lagrangian is given by
\begin{eqnarray}
-{\cal L}&=&-\mu_\Phi^2(\Phi^\dagger\Phi)-\mu_\chi^2(\chi^\dagger \chi)
+\lambda_\Phi(\Phi^\dagger\Phi)^2+\lambda_\chi(\chi^\dagger \chi)^2
+\lambda_4(\Phi^\dagger \chi)(\chi^\dagger \Phi)+M_\Delta^2(\Delta^\dagger\Delta)\nonumber\\
&&+\Big[{C_{ab}\over2}\rho {(\bar \ell_R^c)_{a}} (\ell_{R})_b-\mu\Delta\Phi^\dagger\chi^\dagger+{\kappa\over 2}\rho^*\Delta^2+\lambda \rho^*\Delta \Phi\chi+{\lambda_5\over 2}(\Phi^\dagger\chi)^2+{\rm h.c.}\Big]\;.
\label{Eq_LagrangianDTM}
\end{eqnarray}
%%%%%%%%%%%%%%%
Since  $\chi$ does not couple to the SM fermions due to the $Z_2$ symmetry, the model 
is similar to the Type-I two-Higgs doublet model~\cite{Barger:1989fj}. The doublet $\chi$ can also have a VEV $v_\chi/\sqrt{2}=\langle \chi^0\rangle$ due to the negative mass term of $\chi$. We can define the mixing angle $\sin\gamma=v_\Phi/\sqrt{v_\Phi^2+v_\chi^2}$ to characterize the scalar mixing matrix when only scalar doublets are taken into account, where
$v^2\equiv v_\Phi^2+v_\chi^2+2v_\Delta^2$ with $v=246\,{\rm GeV}$. 
Note that the $\lambda_5$ term in Eq.~(\ref{Eq_LagrangianDTM})
breaks the lepton number symmetry explicitly so that  the
dangerous Majaron can be avoided, 
while sizable values of $\lambda_\Phi$ and $\lambda_\chi$ are needed 
in order to give  the CP even neutral scalar masses and preserve  the stability of the scalar potential. 

The VEV of $\Delta$ in this model is induced by the $\mu$ term in Eq.~(\ref{Eq_LagrangianDTM}), which is proportional to $v_\Phi v_\chi$ 
with fixed values of $M_\Delta$ and $\mu$. In the model, we have two singly-charged physical states $S_1^\pm$ and $S_2^\pm$ besides the unphysical Goldstone boson $G^\pm$, originated from the mixing among $\Phi^\pm$, $\chi^\pm$ and $\Delta^\pm$. 
The values of the mixing elements between the doublets and $\Delta$ are also
proportional to  $v_\Delta$ like MTM. 
Moreover, the term $\lambda\rho^*\Delta\Phi\chi$ and its hermitian conjugate provide 
 another source for the $\rho^{\pm\pm}-\Delta^{\pm\pm}$ mixing apart from the $\kappa$ term, 
with the contribution to $\theta$  approximately proportional to $\sin 2\gamma$. The results on the scalar
mass spectra are given in Appendix~\ref{App_Mass}.2.

The mechanism  for the neutrino mass generation in DTM is similar to that in MTM. But,
 the main coupling related to $\rho^{\pm\pm}$ is from the effective dimension-5 effective operator $\rho (\chi\Phi)^2$. 
 The formula for the neutrino mass matrix is given by
\begin{eqnarray}
(M_\nu)_{\ell\ell'} &=& 
{1\over (16\pi^2)^2}  {2C_{\ell\ell'}m_{\ell}m_{\ell'}\over v^2c_\gamma^2}
\Big[\Big(\mu{s_{2\theta}\over2}A_{(a1)} +{\kappa\over2}A_{(a2)}+{\lambda v\over 2}A_{(a3)}\Big)\nonumber\\
 &&+A_{(b)}+A_{(c)}+A_{(d)}\Big]\;,
\label{eq4}
\end{eqnarray}
where $A_{(ai)}$ and $A_{(j)}$ with $i=1,2$ and $3$ and $j=b,c$ and $d$ are defined in Eqs.~(B12)-(B15), respectively.
In Eq.~(\ref{eq4}), there is a new contribution proportional to $\lambda$, which is of $\mathcal{O}(v_\Delta / v)$.
Note that the elements of
the neutrino mass matrix in MTM  are  of  $\mathcal{O}(v_\Delta^2 / v^2)$.

It is crucial that the above types of the two-loop neutrino mass generation, 
 in which $\bar \ell_R^c\ell_R\rho $ is the only source of 
the LFV,
can lead to an interesting structure for the neutrino mass matrix. The relative sizes among the matrix elements are determined 
by the  combination factors of $C_{\ell\ell'}m_{\ell}m_{\ell'}$. Assuming that each value of $C_{\ell\ell'}$ is at the same order,
there exist interesting hierarchies for the mass matrix elements, given by
\begin{eqnarray}
(M_\nu)_{ee}\ll (M_\nu)_{e\mu} \ll (M_\nu)_{e\tau}\ll (M_\nu)_{\mu\mu}\ll (M_\nu)_{\mu\tau}\ll (M_\nu)_{\tau\tau}\;.
\label{Eq_neutrinomasshierarchy}
\end{eqnarray}
In particular, $(M_{\nu})_{ee}$ is much less than $(M_{\nu})_{\tau\tau}$ due to $m_e^2/m_\tau^2\sim (1/\,3500^2)$.
In Refs.~\cite{Xing:2002ta, Xing:2002ap,Frampton:2002yf,Desai:2002sz, Guo:2002ei, Honda:2003pg},
it has been shown that only the normal hierarchy for the neutrino mass spectrum
can have the matrix textures in which $(M_{\nu})_{ee}$ together with another matrix element is zero.
Clearly, as the mass hierarchies in Eq.~(\ref{Eq_neutrinomasshierarchy}) naturally realize $(M_{\nu})_{ee}\simeq (M_{\nu})_{e\mu}\simeq 0$,
  both MTM and DTM predict the normal neutrino mass hierarchy.

Recall that in the standard parametrization~\cite{Agashe:2014kda,Chau:1984fp}, the neutrino mixing matrix $V_{\rm PMNS}$ is given by
\begin{eqnarray}
V_{\rm PMNS}=\left(\begin{array}{ccc}
c_{12}c_{13}&s_{12}c_{13}&s_{13}e^{-i\delta}\\
-s_{12}c_{23}-c_{12}s_{23}s_{13}e^{i\delta}&c_{12}c_{23}-s_{12}s_{23}s_{13}e^{i\delta}&s_{23}c_{13}\\
s_{12}s_{23}-c_{12}c_{23}s_{13}e^{i\delta}&-c_{12}s_{23}-s_{12}c_{23}s_{13}e^{i\delta}&c_{23}c_{13}\\
\end{array}\right)
\left(\begin{array}{ccc}
1&0&0\\
0&e^{i\alpha_{21}/2}&0\\
0&0&e^{i\alpha_{31}/2}\\
\end{array}
\right)\,,~
\end{eqnarray}
where $s_{ij} (c_{ij}$)=$\sin\theta_{ij}$ ($\cos\theta_{ij}$) with $\theta_{ij}$ being the mixing angles, $\delta$ is the Dirac phase, and 
$\alpha_{21}$ and $\alpha_{31}$ are the two two Majorana phases. 
Note that one of the Majorana phases
 can be  absorbed by the chiral fermion fields if there exists one massless neutrino. 
For  given values of mass square splittings and mixing angles, there are only two solutions for the three CP phases of
  $\delta$, $\alpha_{21}$ and $\alpha_{31}$, along with the lightest neutrino mass $m_0$, to satisfy
 the mass hierarchies in Eq.~(\ref{Eq_neutrinomasshierarchy}). In particular, by using the central values of the global fitting result 
 for the normal hierarchy mass spectrum,  given by~\cite{Agashe:2014kda}
\begin{eqnarray} 
\sin^2\theta_{12}=0.308\pm0.017\;,\;
\sin^2\theta_{23}=0.437^{+0.033}_{-0.023}\;,\;
\sin^2\theta_{13}=0.0234^{+0.0020}_{-0.0019}\;,\;\\
\Delta m_{21}^2=\Big(7.54^{+0.26}_{-0.22}\Big)\times10^{-5}\,{\rm eV}\;,\;\Delta m_{32}^2=(2.43\pm0.06)\times10^{-3}\,{\rm eV},\label{Eq_NeutrinoOscillation}
\end{eqnarray}
we find that 
\begin{eqnarray}
&&(i): m_0=5.14\times10^{-3}\,{\rm eV}\;,\;\delta=0.60\,\pi\;,\;\alpha_{21}=0.11\,\pi\;,\;\alpha_{31}=0.53\,\pi\;,
\label{eq9}
\\
&&(ii): m_0=5.14\times10^{-3}\,{\rm eV}\;,\;\delta=1.40\,\pi\;,\;\alpha_{21}=1.11\,\pi\;,\;\alpha_{31}=1.47\,\pi\;.
\label{eq10}
\end{eqnarray}
Note that both solutions in Eqs.~(\ref{eq9}) and (\ref{eq10}) have the same value for
$m_0$ but  different CP phases.
It is interesting to see that the predicted Dirac phase $\delta=1.40\pi$ in (ii) of Eq.~(\ref{eq10})
agrees well with that given by
  the global fitting result in Ref.~\cite{Agashe:2014kda}.
 Taking (ii) in Eq.~(\ref{eq10})  as the input parameters, the neutrino mass matrix is then given by

\begin{eqnarray}
M_\nu=
\left(\begin{array}{ccc}
0&0&1.04e^{i1.93\pi}\\
0&2.42\,e^{i0.57\pi}&2.32\,e^{i0.50\pi} \\
1.04\,e^{i1.93\pi}&2.32\,e^{i0.50\pi}&2.7\,9\,e^{i0.55\pi}\\
\end{array}
\right)\,,
\label{Eq_Mnmatrix}
\end{eqnarray}
in unit of $10^{-11}$ GeV.
Note that the empty values for $(M_\nu)_{ee}$ and $(M_\nu)_{e\mu}$ can be placed by 
some small non-zero values when any of the parameters in (ii) is under slightly shifting.

\section{Constraints from lepton flavor violation processes and electroweak oblique parameters }
\begin{figure}
\includegraphics[width=16cm]{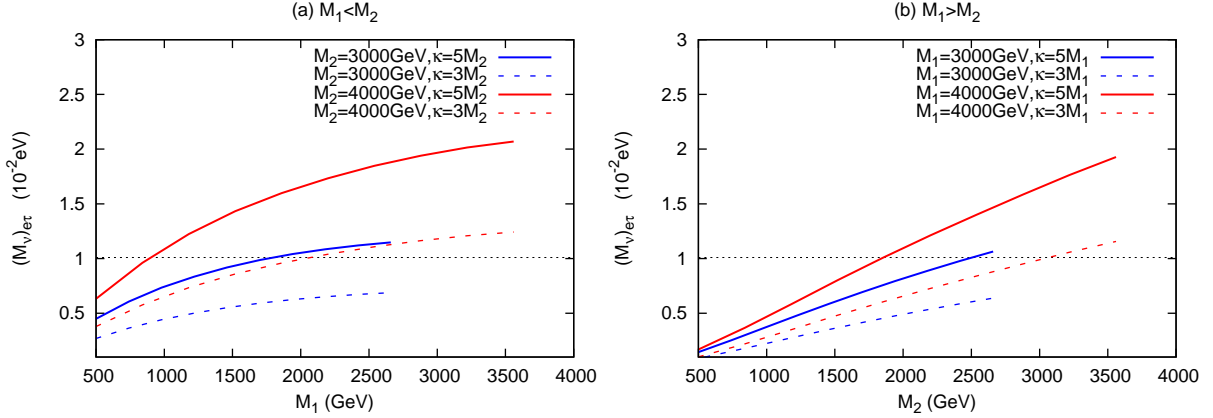} 
\caption{Plots for $(M_\nu)_{e\tau}$ with (a) $M_1<M_2$ and (b) $M_1>M_2$.}
\label{Fig_NeutrinoMet}
\end{figure}

In both MTM and DTM, as the coupling matrix elements $C_{ab}$ are the only sources of the LFV,  
the processes of $\ell\rightarrow \ell'\ell''\ell'''$ ($\ell\rightarrow \ell'\gamma$) with the tree-level (one-loop)
contributions involving $\rho^{\pm\pm}$ could give significant constraints on $C_{ab}$.
%, 
 However, 
those on $C_{ee}$ and $C_{e\mu}$ can be ignored since
 they do not affect the tiny matrix elements $(M_{\nu})_{ee}$ and $(M_{\nu})_{e\mu}$
 when we discuss the neutrino mass spectrum.
 Among the current experimental  bounds, ${\rm Br}(\mu^+\rightarrow e^+\gamma)<5.7\times 10^{-13}$~\cite{Adam:2013mnn} is
 the most stringent one to  limit  of $C_{ab}$. 
 In particular, we can obtain~\cite{Geng:2014gua}
\begin{eqnarray}
|C_{e\tau}|^2\Big({c_\theta^2\over M_1^2}+{s_\theta^2\over M_2^2}\Big)<\Big({0.336\over {\rm TeV}}\Big)^2\;.
\label{Eq_Cet}
\end{eqnarray}
It is obvious that  the largest allowed value of $|C_{e\tau}|_{\rm max}$
from Eq.~(\ref{Eq_Cet})  depends only
on $M_1$ since $s_\theta$ is of order $10^{-2}$. 
To account for the current experimental data on the neutrino masses as obtained in Eq.~(\ref{Eq_Mnmatrix}),
 the matrix element $(M_\nu)_{e\tau}$ should be 
 around $1.04\times 10^{-11}{\rm GeV}$. As a result, we can use this value to check whether the mechanism of the neutrino mass generation can work, as shown in Fig.~\ref{Fig_NeutrinoMet}. 
 The value of $\kappa$ is taken to be $\kappa<{\rm max}(M_1,M_2)$,
 constrained by the perturbativity~\cite{Nebot:2007bc}. In general,
  a larger allowed value of $\kappa$ is more possible to give a correct value of $(M_\nu)_{e\tau}$. 
To obtain the right values for the neutrino masses, at least one of $M_1$ and $M_2$ should roughly larger than $2.5\,{\rm TeV}$. 
In Fig.~\ref{Fig_NeutrinoMet}a, $(M_\nu)_{e\tau}$
 behaves approximately as an increasing function of $M_2$ due to the weak bound on $C_{\ell\ell'}$ from the LFV processes, while 
in Fig.~\ref{Fig_NeutrinoMet}b it is linearly proportional to $M_2$. 

The experimental constraints from the $\mu\rightarrow e$ conversion could also give some hints on $M_1$. 
To illustrate the result, we pick out some of the experimental bounds, given by 
$B_{\mu\rightarrow e}^{\rm Au}<7\times 10^{-13}$~\cite{Bertl:2006up}, $B_{\mu\rightarrow e}^S<7\times 10^{-11}$~\cite{Badertscher:1980bt}, $B_{\mu\rightarrow e}^{\rm Ti}<4.3\times10^{-12}$~\cite{Dohmen:1993mp}, 
and $B_{\mu\rightarrow e}^{\rm Pb}<4.6\times10^{-11}$~\cite{Honecker:1996zf}. For MTM and DTM, the dominant 
contributions come from $\gamma$ and $Z$ penguin diagrams,
which lead to
\begin{eqnarray}
B_{\mu\rightarrow e}^{A}={2G_F m_\mu^5\over \Gamma_A^{\rm capt}}\Big|eA_LD(A)+g_{RV}^{(p)}V^{p}(A)+g_{RV}^{(n)}V^{n}(A)\Big|^2,
\end{eqnarray}
where $\Gamma_A^{\rm capt}$ is the muon capture rate for the nucleus $A$,  the coefficients $A_L$ and $g_{RV}^{(p,n)}$ correspond 
to the dipole and vector contributions, and $D(A)$ and $V^{p,n}(A)$ are the overlapping functions between $e$ and $\mu$ (see Ref.~\cite{Kitano:2002mt} for the details), respectively.
Explicitly, we have
\begin{eqnarray}
A_L={e\sqrt{2}\over 192\pi^2 M_\rho^2 G_F}\sum_l{C_{\mu l}C_{e l}^*}\;,\;
g_{RV(q)}={s_W^2\over 2\pi^2}{5M_W^2\over 9M_\rho^2}Q_q\sum_l{C_{\mu l}C_{e l}^*},
\end{eqnarray}
where $Q_q$ is the electric charge of the quark $q$ and
 $g_{RV(q)}$ is the vector coupling with the quark $q$, mainly  from the $\gamma$ penguin diagram as 
the Z  diagram is suppressed by the charged lepton masses. 
Based on the valence quark model, one has the relations between $g_{RV}^{(n)}$ and $g_{RV}^{(q)}$, given by
  $g_{RV}^{(p)}=2g_{RV(u)}+g_{RV(d)}$ and $g_{RV}^{(n)}=2g_{RV(d)}+g_{RV(u)}$~\cite{Kitano:2002mt}. 
  By taking $M_\rho=1{\rm TeV}$, $C_{e\tau}=0.33$, and $C_{\mu\tau}=0.0033$, 
we find
\begin{eqnarray}
B_{\mu\rightarrow e}^{\rm Au}&=&1.4\times 10^{-14}\;,
B_{\mu\rightarrow e}^{\rm S}=8.3\times 10^{-15}\;,
B_{\mu\rightarrow e}^{\rm Ti}=1.2\times 10^{-14}\;,\nonumber\\
B_{\mu\rightarrow e}^{\rm Pb}&=&9.8\times 10^{-15}\;,
B_{\mu\rightarrow e}^{\rm Al}=7.5\times 10^{-15}\;,
\end{eqnarray}
which satisfy all 
the corresponding 
experimental limits. The improvement on the sensitivity of the $\mu-e$ conversion~\cite{Ref:COMET,Ref:PRISM} in the future will either detect the signal or put some more stringent constraint on the models.

It is interesting to note that the neutrinoless double beta decay in our models can have 
a significant different feature  from other models with radiative neutrino 
mass generations. In MTM and DTM, the short-range contributions to the decay dominate 
over the  traditional long-range ones~\cite{Chen:2006vn, Chen:2007dc},
with the decay amplitudes  proportional to $(M_\nu)_{ee}$. 
 It is clear that the long-range parts can be safely neglected  
 due to the small electron mass in $(M_\nu)_{ee}$,
 whereas  the short-range ones
are  proportional only to the Yukawa coupling $C_{ee}$. 
As a result, by calculating $0\nu\beta\beta$, the upper limit on $|C_{ee}|$ could be derived, despite of the fact that it is 
 ignored when discussing the neutrino mass matrix. The half life for $0\nu\beta\beta$ 
 is given by~\cite{Pas:2000vn, Deppisch:2012nb}
\begin{eqnarray}
T_{1/2}^{0\nu\beta\beta}=(G_{01}|\epsilon_3^{LLL}|^2|\mathcal{M}_3|^2)^{-1}\;,
\end{eqnarray}
which leads to~\cite{Gustafsson:2014vpa}
\begin{eqnarray}
\epsilon_3^{LLL}=m_p\,(2C_{ee}^*s_{2\theta}){v_\Delta\over \sqrt{2}}{M_1^2-M_2^2\over M_1^2M_2^2}\;,
\label{eq19}
\end{eqnarray}
where $G_{01}$ and $|\mathcal{M}_3|$ are the phase space factor and the matrix element for the hadronic sector, respectively,
and $\epsilon_3^{LLL}$ is the coefficient, which is effectively related to the dimension-9 operator $(\bar u_L \gamma_\mu d_L) (\bar u_L \gamma^\mu d_L) (\bar e_R e_R^c)$, defined in Refs.~\cite{Pas:2000vn, Deppisch:2012nb},
and  $m_p$ is the proton mass. Note that the coefficient  in Eq.~(\ref{eq19}) has no explicit dependence on
 the electron mass. 
 If one takes $M_1=1$ and  $M_2=1.5$ TeV in MTM, resulting in the maximal value of mixing
$|\sin2\theta|=0.04$, the upper bounds on $|C_{ee}|$ for different target nuclei can be estimated as
\begin{eqnarray}
|C_{ee}^{\rm (Ge)}|<0.088\;,\;|C_{ee}^{\rm (Xe)}|<0.067\;,\;|C_{ee}^{\rm (Nd)}|<0.36\;,\;\nonumber\\
|C_{ee}^{\rm (Te)}|<0.096\;,\;|C_{ee}^{\rm (Se)}|<0.36\;,\;|C_{ee}^{\rm (Mo)}|<0.13\;,\;
\end{eqnarray}
by comparing with experimental upper limits~\cite{Agostini:2013mzu,KamLANDZen,Gando:2012zm,Argyriades:2008pr,Arnaboldi:2008ds,Arnold:2005rz,Barabash:2010bd}. When including the effect of $\lambda$ in DTM, 
a larger contribution to $\sin 2\theta$ could make these upper bounds on $|C_{ee}|$ to be around $20\%$  lower.

\begin{figure}
\includegraphics[width=16cm]{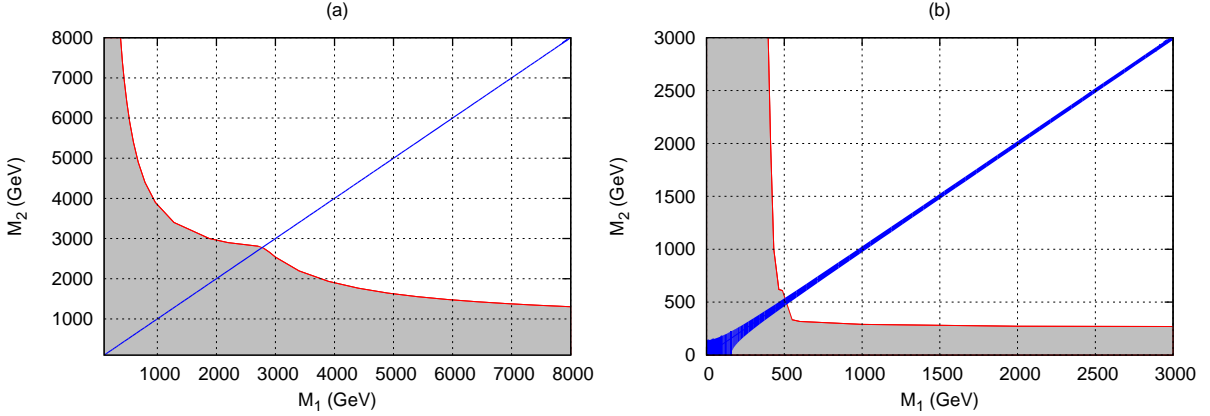} 
\caption{Allowed regions in $M_1-M_2$ plane for (a) MTM and (b) DTM, where
$\kappa=5\,{\rm max}(M_1,M_2)$ and $C_{e\tau}=(C_{e\tau})_{\rm max}$
are used in MTM and DTM,
while  $s_\gamma=0.4$, $\lambda=2$, $\lambda_4=1$ and $\lambda_5=1.5$ are taken in in DTM. 
Gray areas represent regions without
enough neutrino masses, and blue regions located at $M_1\simeq M_2$ are disallowed due to the mixing term between $\rho^{\pm\pm}$ and $\Delta^{\pm\pm}$.}
\label{Fig_NeutrinoM1M2}
\end{figure}

Combing the typical value of $(M_\nu)_{e\tau}$ and the constraints from the LFV processes, at least one of $M_{1}$ and $M_2$ should be heavier than around $2.7\,{\rm TeV}$ in MTM. On the other hand, to get $M_1=\mathcal{O}(10^2)$ GeV, $M_2$ needs to be much larger, 
at least $4$ TeV. However, it is more difficult to get the value of $M_2$ less than $1\,{\rm TeV}$ with a large  $M_1$, which means that to get $M_\rho=\mathcal{O}(100){\rm GeV}$, at least $M_\Delta\gtrsim 4$ TeV is required. 
Consequently,
%%%%%%%%%%%%%
 it is possible to detect the signals of $\rho^{\pm\pm}$,  mainly through the pair production of $\rho^{++}\rho^{--}$ 
 and the subsequently decays with  the same sign charged leptons in the final states at the LHC~\cite{delAguila:2013mia,King:2014uha}. 
 We present the related results in Fig.~\ref{Fig_NeutrinoM1M2}a. Similar conclusions have been also shown in Fig.~2 and 8.8 of Refs.~\cite{delAguila:2011gr, Aparici:2013xga}, respectively, but they allowed some of the region with $M_1\approx M_2\approx 1.5\,{\rm TeV}$, which is forbidden in this paper. 
In DTM, the neutrino masses could be lifted up more easily by using a sizable $\lambda$ as well as $\sin\theta$.  
Moreover,
there is a new contribution to the neutrino masses from the dimension-5 operator $\rho\Phi^2\chi^2$ in DTM instead of
the dimension-7 one $\rho [(D_\mu\Phi)\Phi]^2$ in MTM. As an example, if
we take $s_\gamma=0.4$, $\lambda=2$, $\lambda_4=1$, and $\lambda_5=1.5$, along with the same 
values of $\kappa$ and $C_{e\tau}$ in MTM,
 one finds that $M_2\lesssim 400$ GeV ($\gtrsim 2$ TeV)  
   for $M_1\gtrsim 550$ GeV ($\lesssim 400$ GeV).
   %, 
   The relevant result is displayed in Fig.~\ref{Fig_NeutrinoM1M2}b.

Finally, we briefly discuss the effects of 
the electroweak oblique parameters $S$ and $T$  in our models. 
First of all, in MTM and DTM,
we find that the typical value of $S$ is of  order $10^{-3}$, which is lower than the current experimental sensitivity. 
For the $T$ parameter, the mixing between  $\Delta$ and  $\Phi$ gives a logarithmic divergent $T$ due to the non-unity of $\rho$, but this part could be ignored when $v\ll 5{\rm GeV}$. In this case,
 the main contribution to $T$ is given by the mixing between $\Delta$ and $\rho$, denoted as $T_{(\rho-\Delta)}$. 
 It is basically a negative value whose magnitude is limited due to the small mixing angle $|s_\theta|\lesssim0.02$. 
 For example, taking $M_1=2\,{\rm TeV}$, $M_2=4\,{\rm TeV}$, $v_\Delta=0.5{\rm GeV}$, and $\kappa=5M_2$, 
 it only leads to  $T_{(\rho-\Delta)}=-5\times10^{-5}$. 
 However, in DTM the mixing angle $|s_{\theta}|$ can be enhanced by $\lambda$, which is independent of $v_\Delta$.
Using the input values for the above parameters, and  $\lambda=3$, we  get $T_{(\rho-\Delta)}=-0.001$. 
Meanwhile, the mixing between $\Phi$ and $\chi$ can also provide a sizable value to the corresponding parameter $T_{(\Phi-\chi)}$. 
For example, $T_{(\Phi-\chi)}=0.04$ with $s_\alpha=s_\gamma=0.4$, $\lambda_4=1$, and $\lambda_5=1.5$, which still fulfills the experimental bound $-0.02<\Delta T<0.12$ at $1.5-1.8\,\sigma$ confident level~\cite{Agashe:2014kda}. 
The relevant formulae for $T$  are summarized in Appendix \ref{App_Tpar}.

\section{Conclusions}
We have studied the two models of  MTM and DTM, in which Majorana neutrino masses are
generated radiatively by two-loop diagrams due to the Yukawa 
 $\rho \bar \ell_R^c \ell_R$  and  effective  $\rho^{\pm\pm} W^\mp W^\mp$ couplings.
We have shown that the lepton violating processes, in particular,  $\mu^+\rightarrow e^+\gamma$
can give stringent constraints on the new Yukawa coupling $C_{\ell\ell'}$. 
By combing with the perturbativity condition for the coupling $\kappa$, the light neutrino mass element $(M_\nu)_{e\tau}$ can 
limit the allowed values of $M_1$ and $M_2$. In particular, we have found that only $M_1$ and $M_2$ with ${\rm TeV}$ scale can lead to 
the correct sizes of the neutrino masses.
We have illustrated that the normal neutrino mass hierarchy is a generic feature in these two-loop neutrino mass generation 
 models. Moreover,
 by using the central values of the neutrino oscillation data and comparing with the global fitting result in the 
 literature~\cite{Agashe:2014kda}, we have obtained   
the unique neutrino mass matrix in Eq.~(\ref{Eq_Mnmatrix}) and  predicted the Dirac and two Majorana CP phases to be $1.40\pi$, 
$1.11\pi$ and $1.47\pi$, respectively. 
Finally, we emphasize that the neutrinoless double beta decays can be very large as they are dominated by the short-distance contributions
at tree-level, which can be tested in the future experiments and used to constrain the element of $C_{ee}$.

\begin{acknowledgments}
This work was supported in part by National Center for Theoretical Sciences, National Science
Council (Grant No. NSC-101-2112-M-007-006-MY3) and National Tsing Hua University (Grant No. 104N2724E1).
\end{acknowledgments}
\appendix
\section{Scalar mass spectra in MTM and DTM}\label{App_Mass}
\subsection{MTM}
The non-self-Hermitian terms in Eq.~(\ref{Eq_TripletLagrangian}) can be expanded as fellows:
\begin{eqnarray}
\Delta (\Phi^\dagger)^2&=&\Delta^{++}{(\Phi^+)^*}^2+2{\Delta^{+}\over\sqrt{2}}(\Phi^+)^*(\Phi^0)^*
+\Delta^0{(\Phi^0)^*}^2\;,
\\
\rho^*\Delta^2&=&2\rho^*\Big[\Delta^{++}\Delta^{0}-{1\over2}({\Delta^+})^2\Big]\;.
\end{eqnarray}
After obtaining the explicit forms of the scalar potential, we can write down its tadpole conditions
\begin{eqnarray}
&&-\mu_\Phi^2 v_\Phi+\lambda_\Phi v_\Phi^3-\sqrt{2}\mu v_\Phi v_\Delta=0\;,\\
&&M_\Delta^2 v_\Delta -\mu{v_\Phi^2\over \sqrt{2}}=0\;,
\end{eqnarray}
which give
\begin{eqnarray}
v_\Delta={\mu\over\sqrt{2}}{v_\Phi^2\over M_\Delta^2}\;.
\end{eqnarray}
The mixing matrices of CP odd neutral  and singly charged states are written as
\begin{eqnarray}
M_I^2=\left(\begin{array}{cc}
t_\beta&-t_\beta \;\;\;\\
-t_\beta&1\;\;\;\\
\end{array}\right)M_\Delta^2\;,~~~
M_{\pm}^2=\left(\begin{array}{cc}
t_\beta'^2&-t_\beta'\;\;\;\\
-t_\beta'&1\;\;\;\\
\end{array}\right)M_\Delta^2\;,
\end{eqnarray}
where $t_\beta=2v_\Delta/v_\Phi$ and $t_\beta'=\sqrt{2}v_\Delta/v_\Phi$. 
The masses of the neutral CP odd state $A$ and singly charged states $S^\pm$ are given by
 $M_A=M_\Delta/c_\beta^2$ and   $M_S=M_\Delta/c_\beta'\,^2$, respectively.

The doubly-charge mixing matrix is given by
\begin{eqnarray}
\left(\begin{array}{cc}
M_\rho^2&M_{12}^{\pm\pm}\;\;\;\\
M_{12}^{\pm\pm}&M_\Delta^2\;\;\;\\
\end{array}\right)\;,
\label{A7}
\end{eqnarray}
where 
\begin{eqnarray}
M_{12}^{\pm\pm}={\kappa v_\Delta\over\sqrt{2}}\;.
\end{eqnarray}
One can easily diagonalize Eq.~(\ref{A7}), leading to 
 the mixing angle $\theta$ 
\begin{eqnarray}
t_{2\theta}={2M_{12}^{\pm\pm}\over M_\rho^2-M_\Delta^2}\;,
\end{eqnarray}
and  eigenvalues of the eigenstates $P_{1,2}$ 
\begin{eqnarray}
M_{1}^2&=&M_\rho^2-{s_\theta^2 \over 1-2s_\theta^2}(M_\Delta^2-M_\rho^2)\;,\;\nonumber\\
M_{2}^2&=&M_\Delta^2+{s_\theta^2 \over 1-2s_\theta^2}(M_\Delta^2-M_\rho^2)\;.
\end{eqnarray}

\subsection{DTM}
The related operators with $\chi$ in Eq.~(\ref{Eq_LagrangianDTM}) can be written as
\begin{eqnarray}
\Delta \Phi^\dagger\chi^\dagger&=&\Delta^{++}\Phi^{+*}\chi^{+*}
+{\Delta^+\over\sqrt{2}}(\Phi^{+*}\chi^{0*}+\Phi^{0*}\chi^{+*})+\Delta^0\Phi^{0*}\chi^{0*}\;,\nonumber\\
\rho^*\Delta \Phi\chi&=& \rho^{--}\Big[\Delta^{++}\Phi^0\chi^0-
{\Delta^+\over \sqrt{2}}(\Phi^+\chi^0+\chi^+\Phi^0)+\Delta^{0}\Phi^+\chi^+\;\Big]\;.
\end{eqnarray}
The minimization conditions are given by
\begin{eqnarray}
&&-\mu_\Phi^2 v_\Phi+\lambda_\Phi v_\Phi^3-{1\over2}(\lambda_4+\lambda_5)v_\Phi v_\chi^2-{\mu\over\sqrt{2}} v_\Delta v_\chi=0\;,\\
&&-\mu_\chi^2 v_\chi+\lambda_\chi v_\chi^3-{1\over2}(\lambda_4+\lambda_5)v_\chi v_\Phi^2-{\mu\over\sqrt{2}} v_\Delta v_\Phi=0\;,\\
&&M_\Delta^2 v_\Delta -\mu{v_\Phi v_\chi\over \sqrt{2}}=0\;.
\end{eqnarray}
It is convenient to define $\bar v=\sqrt{v_\Phi^2+v_\chi^2}$, and $s_\gamma=v_\chi/\bar v$. The singly charged and CP odd neutral mass matrices are both $3\times 3$. In the diagonalization,
we use the relation $s_\gamma\gg s_\beta'$.
The transformation matrices, $V_I$ and  $V_\pm$, of CP odd neutral  
and  singly charged states can be presented by a set of small quantities $\epsilon_{ij}$ and $\epsilon_{ij}'$,  given as
\begin{eqnarray}
V_I=\left(\begin{array}{ccc}
c_\gamma&-s_\gamma&\epsilon_{13} \;\;\;\\
s_\gamma&c_\gamma&\epsilon_{23}\;\;\;\\
\epsilon_{31}&\epsilon_{32}&1\;\;\;\\
\end{array}\right)\;,
V_{\pm}=\left(\begin{array}{ccc}
c_\gamma&-s_\gamma&\epsilon_{13}' \;\;\;\\
s_\gamma&c_\gamma&\epsilon_{23}'\;\;\;\\
\epsilon_{31}'&\epsilon_{32}'&1\;\;\;\\
\end{array}\right)\;,
\end{eqnarray}
where
\begin{eqnarray}
&&\epsilon_{32}={M_\Delta^2\over M_\Delta^2-\lambda_5\bar v^2}{c_{2\gamma}\over s_{2\gamma}}t_\beta\;,\\
&&\epsilon_{31}=t_\beta\;,\\
&&\epsilon_{13}=(-c_\gamma+{c_{2\gamma}\over 2c_\gamma}{M_\Delta^2\over M_\Delta^2-\lambda_5 \bar v^2})t_\beta\;,\\
&&\epsilon_{23}=(-s_\gamma-{c_{2\gamma}\over 2s_\gamma}{M_\Delta^2\over M_\Delta^2-\lambda_5 \bar v^2})t_\beta\;,
\end{eqnarray}
and
\begin{eqnarray}
&&\epsilon_{32}'={M_\Delta^2\over M_\Delta^2-{1\over2}(\lambda_4+\lambda_5)\bar v^2}{c_{2\gamma}\over s_{2\gamma}}t_\beta'\;,\\
&&\epsilon_{31}'=t_\beta'\;,\\
&&\epsilon_{13}'=\Big(-c_\gamma+{c_{2\gamma}\over 2c_\gamma}{M_\Delta^2\over M_\Delta^2-{1\over2}(\lambda_4+\lambda_5) \bar v^2}\Big)t_\beta'\;,\\
&&\epsilon_{23}'=\Big(-s_\gamma-{c_{2\gamma}\over 2s_\gamma}{M_\Delta^2\over M_\Delta^2-{1\over2}(\lambda_4+\lambda_5) \bar v^2}\Big)t_\beta'\;.
\end{eqnarray}
The mass eigenvalues for two CP odd states are $M_{A_1}^2=\lambda_5 \bar v^2$ and $M_{A_2}^2=M_\Delta^2$, 
while those  for singly charged states 
$M_{S_1}^2={1\over2}(\lambda_4+\lambda_5) \bar v^2$ and $M_{S_2}^2=M_\Delta^2$. 
Finally, the mixing angle between the doubly-charged states is given by
\begin{eqnarray}
t_{2\theta}={\sqrt{2}\kappa v_\Delta+\lambda c_\gamma s_\gamma v_\Phi^2\over M_\rho^2-M_\Delta^2}\;.
\end{eqnarray}
For the mixing of  CP even neutral states, one can focus on the $2\times2$ mixing matrix between $\Phi$ and $\chi$, 
 given by
\begin{eqnarray}
\left(\begin{array}{cc}
2\lambda_\Phi v_\Phi^2&-(\lambda_4+\lambda_5)v_\Phi v_\chi\;\;\;\\
-(\lambda_4+\lambda_5)v_\Phi v_\chi&2\lambda_\chi v_\chi^2\;\;\;\\
\end{array}\right)\;.
\end{eqnarray}
Rotating the states $\Phi_R=c_\alpha h-s_\alpha H$ and $\chi_R=s_\alpha h+c_\alpha H$, we obtain 
 two mass eigenvalues $M_h$ and $M_H$ to be
\begin{eqnarray}
M_h^2&=&(2\lambda_\Phi -(\lambda_4+\lambda_5)t_\gamma t_\alpha)v_\Phi^2c_\beta'^2 c_\gamma^2\;,\\
M_H^2&=&(2\lambda_\Phi +(\lambda_4+\lambda_5){t_\gamma \over t_\alpha})v_\Phi^2c_\beta'^2 c_\gamma^2\;.
\end{eqnarray}
If one takes $m_h=125.7{\rm GeV}$~\cite{Agashe:2014kda} as the state found at the LHC~\cite{atlas:2012gk,cms:2012gu},  
$M_H$ can be obtained as a function of $\lambda_4+\lambda_5$ and $\alpha$.

\section{Two-Loop Neutrino mass}\label{App_neutrinomass}
\subsection{MTM}
We calculate the neutrino masses   by using the t'Hooft-Feynman Gauge. The relevant vertices corresponding to the top vertex in 
Fig.~\ref{Fig_NeutrinoMass}a are $-\mu \Delta^{--}{(\Phi^-)^*}^2$ and $-{\kappa \over 2}{\rho^{++}}^*(\Delta^+)^2$, which yield
\begin{eqnarray}
-i\mathcal{M}^{(a)}_{\ell\ell'}
&=&{iv_\Delta\over (16\pi^2)^2}\Big({C_{\ell\ell'}m_{\ell} m_{\ell'}\over v^2c_{\beta'}^2}\Big)\Big[
-4\Big({\kappa v_\Delta\over v^2}\Big){M_0^2\over M_1^2-M_2^2} (I_{111}-I_{112})\nonumber\\
&&-4{\kappa v_\Delta\over v^2}\Big(c_\theta^2(I_{111}-2I_{121}+I_{221})
+s_\theta^2(I_{112}-2I_{212}+I_{222})\Big)
%&&+{\lambda_4s_{2\theta}\over\sqrt{2} }(I_{111}-I_{121}-I_{112}+I_{122})\nonumber\\
%&&2\sqrt{2}\lambda \Big(c_\theta^2(-I_{111}+2I_{121})+s_\theta^2(-I_{112}+2I_{122})\Big)
\Big](\bar \nu_L^c\nu_L)\;,
\end{eqnarray}
with
\begin{eqnarray}
I_{ijn}=\int_0^1dy_2\int_0^{1-y_1}dy_1\int_0^{1}dx_2\int_0^{1-x_2}dx_1 {2\over 1-x_1}\log(y_1[x_1M_{n}^2+x_2M_{i}'\,^2]+y_2x_1(1-x_1)M_{j}'\,^2),~
\nonumber\\ \label{Eq_Iijk}
\end{eqnarray}
where we can define $M_1'=M_W$ and $M_2'=M_S$. For Fig.~\ref{Fig_NeutrinoMass}b,  we get
\begin{eqnarray}
-i\mathcal{M}^{(b)}_{\ell\ell'}&=&{i\over (16\pi^2)^2}\Big({\sqrt{2}C_{\ell\ell'}m_\ell m_{\ell'}\over v_\Phi}\Big){\sqrt{2}g_W^2\over2}
{s_{2\theta}\over2}{s_{2\beta'}\over2}(M_1^2-M_2^2)(I_{W}^{(b)}-I_{S}^{(b)})(\bar \nu_L^c\nu_L)\;,\nonumber\\
\end{eqnarray}
with
\begin{eqnarray}
I_i^{(b)}&=&\int_0^1dy_2\int_0^{1-y_1}dy_1\int_0^{1}dx_3\int_0^{1-x_3}dx_2\int_0^{1-x_2-x_3}dx_1 {-2y_1(2-x_1-x_2)\over (x_1+x_2)(1-x_1-x_2) m_{i}^2}\;,\nonumber\\
\end{eqnarray}
where
\begin{eqnarray}
m^2_{i}&=&y_1(x_1M_1^2+x_2M_2^2+x_3 M_W^2)+y_2(x_1+x_2)(1-x_1-x_2)M_{i}'\,^2\;.
\end{eqnarray}
The amplitude $\mathcal{M}^{(c)}_{ll'}$ should be equal to $\mathcal{M}^{(b)}_{ll'}$. 
For Fig.~\ref{Fig_NeutrinoMass}d, we have
\begin{eqnarray}
-i\mathcal{M}^{(d)}_{\ell\ell'}&=&{-i\over (16\pi^2)^2}{2g_W^4v_\Delta \over 4\sqrt{2}}s_{2\theta}(m_\ell m_{\ell'}C_{\ell\ell'})
[I^{(d)}(M_{P1}^2)-I^{(d)}(M_{P2}^2)](\bar \nu_L^c\nu_L)\;,
\end{eqnarray}
\begin{eqnarray}
I^{(d)}&=&\int_0^1dy_2\int_0^{1-y_1}dy_1\int_0^{1}dx_2\int_0^{1-x_2}dx_1 {-4\over m_{(d)}^2}\;,
\end{eqnarray}
where
\begin{eqnarray}
m_{(d)}^2=y_1(x_1 M_i^2+x_2 M_W^2)+y_2x_1(1-x_1)M_W^2\;.
\end{eqnarray}
In summary, $A_{(a)}$, $A_{(b)}$, $A_{(c)}$, and $A_{(d)}$ in Eq.~(\ref{Eq_neutrinomassMTM}) are listed as follows:
\begin{eqnarray}
A_{(a)}&=&s_{\beta'}^2\kappa{M_\Delta^2\over M_1^2-M_2^2}\Big[(I_{111}-I_{112})
+{s_\beta'^2\over c_\beta'^2}(I_{121}-I_{122}+I_{211}-I_{212})\nonumber\\
&&{s_\beta'^4\over c_\beta'^4}(I_{221}-I_{222})\Big]+s_{\beta'}^2\kappa\Big[c_\theta^2(I_{111}-I_{121}-I_{211}+I_{221})\nonumber\\
&&+s_\theta^2(I_{112}-I_{122}-I_{212}+I_{222})\Big]\;,\\
A_{(b)}&=&A_{(c)}=-2s_{\beta'}^2\kappa M_W^2(I_W^{(b)}-I_S^{(b)})\;,\\
A_{(d)}&=&2s_{\beta'}^2\kappa{M_W^4\over M_1^2-M_2^2}I^{(d)}\;.
\end{eqnarray}

\subsection{DTM}
If we expand the amplitudes $A_{(a1)}$, $A_{(a2)}$, and $A_{(a3)}$ up to $\mathcal{O}(v_\Delta/v)$, then the results are given by
\begin{eqnarray}
A_{(a1)}&=&0\;,\;A_{(a2)}=0\;,\\
A_{(a3)}&=&s_\gamma c_\theta^2\Big[2c_\gamma^3\epsilon_{31}'I_{111}+2c_\gamma s_\gamma^2 \epsilon_{31}'(I_{121}+I_{211})
+2c_\gamma^2\epsilon_{13}'(I_{131}+I_{311})-2\epsilon_{32}'s_\gamma^3 I_{221}\nonumber\\
&&+2s_\gamma^2\epsilon_{13}'(I_{231}+I_{321})\Big]+c_\gamma c_\theta^2\Big[2c_\gamma^2 s_\gamma \epsilon_{31}' I_{111}-2c_\gamma s_\gamma (\epsilon_{31}' c_\gamma +\epsilon_{32}' s_\gamma)(I_{121}+I_{211})\nonumber\\
&&+2c_\gamma s_\gamma \epsilon_{13}'(I_{131}+I_{311})+2c_\gamma s_\gamma^2 \epsilon_{32}' I_{221}-2c_\gamma s_\gamma \epsilon_{13}'(I_{231}+I_{321})\Big]\nonumber\\
&&-s_\beta' c_\theta^2\Big[2 c_\gamma^3 s_\gamma I_{111}-2c_\gamma s_\gamma (c_\gamma^2 -s_\gamma^2)(I_{121}+I_{211})
-2c_\gamma s_\gamma^3 I_{221}\Big]\;,\\
A_{(b)}&=&A_{(c)}=-2s_{2\theta}{M_1^2-M_2^2\over v} M_W^2(\epsilon_{31}'I_W^{(b)}-s_\gamma \epsilon_{32}'I_{S_1}^{(b)}+\epsilon_{13}'I_{S_2}^{(b)})\;,\\
A_{(d)}&=&2s_{\beta'}{s_{2\theta}\over v}M_W^4I^{(d)}\;.
\end{eqnarray}
where we have used the same notations as those in
Appendix~\ref{App_neutrinomass}1 except
%$I_{ijk}$ is defined in Eq.~(\ref{Eq_Iijk}), but $M_1'=M_W$, 
$M_2'=M_{S_1}$ and $M_3'=M_{S_2}$.

\section{T parameters in MTM and DTM}\label{App_Tpar}

The $T$ parameter due to the mixing between $\rho$ and $\Delta$ in MTM is given by

\begin{eqnarray}
T_{(\rho-\Delta)}&=&{1\over 4\pi s_W^2 M_W^2}\Big[
s_\theta^2 F(M_{1}^2,M_{S^\pm}^2)+c_\theta^2 F(M_{2}^2,M_{S^\pm}^2)-F(M_\Delta^2,M_{S^\pm}^2)\nonumber\\
&&-2c_\theta^2s_\theta^2F(M_{1}^2,M_{2}^2) \Big]\;.\label{Eq_TrhoDelta}
\end{eqnarray}
The above result in Eq.~(\ref{Eq_TrhoDelta}) can also be applied to DTM with replacing $S^\pm$  by $S_2^\pm$.
In DTM, since the contribution from the mixing between $\Phi$ and $\chi$ should also be considered, 
we find
\begin{eqnarray}
T_{(\Phi-\chi)}&=&{1\over 16\pi s_W^2 M_W^2}\Big[(c_{\alpha-\gamma}^2-1)(G(M_{h}^2,M_{W}^2)-G(M_{h}^2,M_{Z}^2))
+s_{\alpha-\gamma}^2 (G(M_{H}^2,M_{W}^2)-G(M_{H}^2,M_{Z}^2))\nonumber\\
&&+s_{\alpha-\gamma}^2(F(M_{h}^2,M_{S_1}^2)-F(M_{h}^2,M_{A_1}^2))
+c_{\alpha-\gamma}^2 (F(M_{H}^2,M_{S_1}^2)-F(M_{H}^2,M_{A_1}^2))\nonumber\\ 
&&+F(M_{S_1}^2,M_{A_1}^2)\Big]\;,
\end{eqnarray}
where the functions $F$, $K$, and $G$ are defined by
\begin{eqnarray}
G(x,y)&=&F(x,y)+4yK(x,y)\;,\;\\
F(x,y)&=&{x+y\over2}-{xy\over x-y}\log{x\over y}\;,\;
K(x,y)={x\log x-y\log y\over x-y}\;.\;
\end{eqnarray}


\begin{thebibliography}{01}
%%%Neutrino Oscillations
\bibitem{Anselmann:1992kc} 
  P.~Anselmann {\it et al.}  [GALLEX Collaboration],
  %``Implications of the GALLEX determination of the solar neutrino flux.,''
  Phys.\ Lett.\ B {\bf 285}, 390 (1992).
\bibitem{Fukuda:1998mi} 
  Y.~Fukuda {\it et al.}  [Super-Kamiokande Collaboration],
  %``Evidence for oscillation of atmospheric neutrinos,''
  Phys.\ Rev.\ Lett.\  {\bf 81}, 1562 (1998)
  [hep-ex/9807003].
\bibitem{Ahmad:2002jz} 
  Q.~R.~Ahmad {\it et al.}  [SNO Collaboration],
  %``Direct evidence for neutrino flavor transformation from neutral current interactions in the Sudbury Neutrino Observatory,''
  Phys.\ Rev.\ Lett.\  {\bf 89}, 011301 (2002)
  [nucl-ex/0204008].
\bibitem{Ahmad:2002ka} 
  Q.~R.~Ahmad {\it et al.}  [SNO Collaboration],
  %``Measurement of day and night neutrino energy spectra at SNO and constraints on neutrino mixing parameters,''
  Phys.\ Rev.\ Lett.\  {\bf 89}, 011302 (2002)
  [nucl-ex/0204009].  
\bibitem{Ahn:2006zza} 
  M.~H.~Ahn {\it et al.}  [K2K Collaboration],
  %``Measurement of Neutrino Oscillation by the K2K Experiment,''
  Phys.\ Rev.\ D {\bf 74}, 072003 (2006)
  [hep-ex/0606032].
%%%%%%%%%%%%%%%%%Type I
\bibitem{TypeIseesaw1}
%\bibitem{Minkowski:1977sc}
  P.~Minkowski,
  %``Mu $\to$ E Gamma At A Rate Of One Out Of 1-Billion Muon Decays?,''
  Phys.\ Lett.\  B {\bf 67}, 421 (1977).
\bibitem{TypeIseesaw2}
T.~Yanagida, in {\it Proceedings of the Workshop on the Unified Theory and the Baryon Number in
the Universe}, edited by O.~Sawada and A.~Sugamoto (KEK, Tsukuba, 1979), p.~95.
\bibitem{TypeIseesaw3}
M.~Gell-Mann, P.~Ramond, and R.~Slansky, in {\it Supergravity},
edited by P.~van Nieuwenhuizen and D.~Freedman (North-Holland, Amsterdam, 1979), p.~315.
\bibitem{TypeIseesaw4}
S.~L.~Glashow, in {\it Proceedings of the 1979 Cargese Summer Institute on Quarks and Leptons},
edited by M.~Levy {\it et al}. (Plenum Press, New York, 1980), p. 687.
\bibitem{TypeIseesaw5}
%\bibitem{Mohapatra:1979ia}
  R.~N.~Mohapatra and G.~Senjanovic,
  %``Neutrino mass and spontaneous parity nonconservation,''
  Phys.\ Rev.\ Lett.\  {\bf 44}, 912 (1980).

%%%%%%%%%%%%%%%%%%TYPE II SEESAW
\bibitem{typeIIseesaw1}
M.~Magg and C.~Wetterich,
  %``Neutrino Mass Problem And Gauge Hierarchy,''
  Phys.\ Lett.\  B {\bf 94}, 61 (1980).
\bibitem{typeIIseesaw2}
   J.~Schechter and J.~W.~F.~Valle,
  %``Neutrino Masses In SU(2) X U(1) Theories,''
  Phys.\ Rev.\  D {\bf 22}, 2227 (1980).
\bibitem{typeIIseesaw3}  
   T.~P.~Cheng and L.~F.~Li,
  %``Neutrino Masses, Mixings And Oscillations In SU(2) X U(1) Models Of
  %Electroweak Interactions,''
  Phys.\ Rev.\  D {\bf 22}, 2860 (1980). 
\bibitem{typeIIseesaw4}
G.~B.~Gelmini and M.~Roncadelli,
  %``Left-Handed Neutrino Mass Scale And Spontaneously Broken Lepton Number,''
  Phys.\ Lett.\  B {\bf 99}, 411 (1981).
\bibitem{typeIIseesaw5}  
G.~Lazarides, Q.~Shafi and C.~Wetterich,
  %``Proton Lifetime And Fermion Masses In An SO(10) Model,''
  Nucl.\ Phys.\  B {\bf 181}, 287 (1981).
\bibitem{typeIIseesaw6}  
  R.~N.~Mohapatra and G.~Senjanovic,
  %``Neutrino Masses And Mixings In Gauge Models With Spontaneous Parity
  %Violation,''
  Phys.\ Rev.\  D {\bf 23}, 165 (1981).
\bibitem{typeIIseesaw7}  
  J.~Schechter and J.~W.~F.~Valle,
  %``Neutrino Decay And Spontaneous Violation Of Lepton Number,''
  Phys.\ Rev.\  D {\bf 25}, 774 (1982).  

%%%%%%%%%%%%%%%%%%%  type III
\bibitem{Foot:1988aq} 
  R.~Foot, H.~Lew, X.~G.~He and G.~C.~Joshi,
  %``Seesaw Neutrino Masses Induced by a Triplet of Leptons,''
  Z.\ Phys.\ C {\bf 44}, 441 (1989).
  %%CITATION = ZEPYA,C44,441;%%
%%%%%%%%%%%%%%%%%%%%%%%
\bibitem{Zee:1980ai} 
  A.~Zee,
  %``A Theory of Lepton Number Violation, Neutrino Majorana Mass, and Oscillation,''
  Phys.\ Lett.\ B {\bf 93}, 389 (1980)
  [Erratum-ibid.\ B {\bf 95}, 461 (1980)].

\bibitem{Ma:2006km} 
  E.~Ma,
  %``Verifiable radiative seesaw mechanism of neutrino mass and dark matter,''
  Phys.\ Rev.\ D {\bf 73}, 077301 (2006)
  [hep-ph/0601225].  

     
%%%%%%%%%%%%%%%%%%%%%%%%%%%%%%%%Two  Loop  
\bibitem{Zee:1985id} 
  A.~Zee,
  %``Quantum Numbers of Majorana Neutrino Masses,''
  Nucl.\ Phys.\ B {\bf 264}, 99 (1986).
  %%CITATION = NUPHA,B264,99;%%
\bibitem{Babu:1988ki} 
  K.~S.~Babu,
  %``Model of 'Calculable' Majorana Neutrino Masses,''
  Phys.\ Lett.\ B {\bf 203}, 132 (1988).
  %%CITATION = PHLTA,B203,132;%%
%\bibitem{Ma:2013mga} 
%  E.~Ma,
  %``Radiative Origin of All Quark and Lepton Masses through Dark Matter with Flavor Symmetry,''
%  Phys.\ Rev.\ Lett.\  {\bf 112}, no. 9, 091801 (2014)
%  [arXiv:1311.3213 [hep-ph]].
%\bibitem{Okada:2014vla} 
%  H.~Okada,
  %``Two loop Induced Dirac Neutrino Model and Dark Matters with Global $U(1)'$ Symmetry,''
%  arXiv:1404.0280 [hep-ph].
  %%CITATION = ARXIV:1404.0280;%%  
%\bibitem{Kanemura:2014rpa} 
%  S.~Kanemura, T.~Matsui and H.~Sugiyama,
  %``Neutrino Mass and Dark Matter from Gauged $U(1)_{B-L}$ Breaking,''
%  Phys.\ Rev.\ D {\bf 90}, no. 1, 013001 (2014)
%  [arXiv:1405.1935 [hep-ph]].  
%\bibitem{Okada:2014qsa} 
%  H.~Okada, T.~Toma and K.~Yagyu,
  %``Inert Extension of the Zee-Babu Model,''
%  Phys.\ Rev.\ D {\bf 90}, no. 9, 095005 (2014)
%  [arXiv:1408.0961 [hep-ph]].  

%%%%%%%%%%%%%effective operator%%%%%%%%%%%%%%%%%%
\bibitem{Chen:2006vn} 
  C.~S.~Chen, C.~Q.~Geng and J.~N.~Ng,
  %``Unconventional Neutrino Mass Generation, Neutrinoless Double Beta Decays, and Collider Phenomenology,''
  Phys.\ Rev.\ D {\bf 75}, 053004 (2007)
  [hep-ph/0610118].
  
  \bibitem{Chen:2007dc}
  C.~S.~Chen, C.~Q.~Geng, J.~N.~Ng and J.~M.~S.~Wu,
  %``Testing radiative neutrino mass generation at the LHC,''
  JHEP {\bf 0708}, 022 (2007)
  [arXiv:0706.1964 [hep-ph]].

   
%%%%%%%%%%%%%%%%%%%% Three Loop  %%%%%%%%%%%%%%%%%%
\bibitem{Krauss:2002px} 
  L.~M.~Krauss, S.~Nasri and M.~Trodden,
  %``A Model for neutrino masses and dark matter,''
  Phys.\ Rev.\ D {\bf 67}, 085002 (2003)
  [hep-ph/0210389].
  %%CITATION = HEP-PH/0210389;%%
\bibitem{Aoki:2008av} 
  M.~Aoki, S.~Kanemura and O.~Seto,
  %``Neutrino mass, Dark Matter and Baryon Asymmetry via TeV-Scale Physics without Fine-Tuning,''
  Phys.\ Rev.\ Lett.\  {\bf 102}, 051805 (2009)
  [arXiv:0807.0361 [hep-ph]].
  %%CITATION = ARXIV:0807.0361;%%
\bibitem{Gustafsson:2012vj} 
  M.~Gustafsson, J.~M.~No and M.~A.~Rivera,
  %``Predictive Model for Radiatively Induced Neutrino Masses and Mixings with Dark Matter,''
  Phys.\ Rev.\ Lett.\  {\bf 110}, no. 21, 211802 (2013)
  [Erratum-ibid.\  {\bf 112}, no. 25, 259902 (2014)]
  [arXiv:1212.4806 [hep-ph]].
%\bibitem{Kajiyama:2013lja} 
%  Y.~Kajiyama, H.~Okada and K.~Yagyu,
  %``$T_7$ Flavor Model in Three Loop Seesaw and Higgs Phenomenology,''
%  JHEP {\bf 1310}, 196 (2013)
%  [arXiv:1307.0480 [hep-ph]].
%\bibitem{Ahriche:2014cda} 
%  A.~Ahriche, C.~S.~Chen, K.~L.~McDonald and S.~Nasri,
  %``A Three-Loop Model of Neutrino Mass with Dark Matter,''
%  Phys.\ Rev.\ D {\bf 90}, no. 1, 015024 (2014)
%  [arXiv:1404.2696 [hep-ph]].  
%\bibitem{Chen:2014ska} 
%  C.~S.~Chen, K.~L.~McDonald and S.~Nasri,
  %``A Class of Three-Loop Models with Neutrino Mass and Dark Matter,''
%  Phys.\ Lett.\ B {\bf 734}, 388 (2014)
%  [arXiv:1404.6033 [hep-ph]].  
%\bibitem{Hatanaka:2014tba} 
%  H.~Hatanaka, K.~Nishiwaki, H.~Okada and Y.~Orikasa,
  %``A Three-Loop Neutrino Model with Global $U(1)$ Symmetry,''
%  arXiv:1412.8664 [hep-ph].  
%%%%%%%%%%%%%%%%%%%%%%%%%%%%%%%%%%%%%%%%%%%%%%%%%%%%%%%%%%%
\bibitem{Geng:2014gua} 
  C.~Q.~Geng, D.~Huang and L.~H.~Tsai,
  %``Loop-induced Neutrino Masses: A Case Study,''
  Phys.\ Rev.\ D {\bf 90}, 113005 (2014)
  [arXiv:1410.7606 [hep-ph]]. 
 


\bibitem{Gustafsson:2014vpa} 
  M.~Gustafsson, J.~M.~No and M.~A.~Rivera,
  %``Radiative neutrino mass generation linked to neutrino mixing and $0\nu\beta\beta$-decay predictions,''
  Phys.\ Rev.\ D {\bf 90}, 013012 (2014)
  [arXiv:1402.0515 [hep-ph]].

\bibitem{King:2014uha} 
  S.~F.~King, A.~Merle and L.~Panizzi,
  %``Effective theory of a doubly charged singlet scalar: complementarity of neutrino physics and the LHC,''
  JHEP {\bf 1411}, 124 (2014)
  [arXiv:1406.4137 [hep-ph]].
  

\bibitem{Sierra:2014rxa} 
  D.~A.~Sierra, A.~Degee, L.~Dorame and M.~Hirsch,
  %``Systematic classification of two-loop realizations of the Weinberg operator,''
  arXiv:1411.7038 [hep-ph].


\bibitem{Chen:2010ir} 
  C.~S.~Chen and C.~Q.~Geng,
  %``Relating Neutrino Masses by dilepton modes of Doubly Charged Scalars,''
  Phys.\ Rev.\ D {\bf 82}, 105004 (2010)
  [arXiv:1005.2817 [hep-ph]].  
 
  
\bibitem{delAguila:2011gr} 
  F.~del Aguila, A.~Aparici, S.~Bhattacharya, A.~Santamaria and J.~Wudka,
  %``A realistic model of neutrino masses with a large neutrinoless double beta decay rate,''
  JHEP {\bf 1205}, 133 (2012)
  [arXiv:1111.6960 [hep-ph]].

%%%%%%0nbb%%%
\bibitem{delAguila:2012nu} 
  F.~del Aguila, A.~Aparici, S.~Bhattacharya, A.~Santamaria and J.~Wudka,
  %``Effective Lagrangian approach to neutrinoless double beta decay and neutrino masses,''
  JHEP {\bf 1206}, 146 (2012)
  [arXiv:1204.5986 [hep-ph]].
  
%%%%%%%%%%%%%%%%%%%%%%  PDG%%%
\bibitem{Agashe:2014kda} 
  K.~A.~Olive {\it et al.}  [Particle Data Group Collaboration],
  %``Review of Particle Physics,''
  Chin.\ Phys.\ C {\bf 38}, 090001 (2014).  
  
  \bibitem{Chen:2012vm}
  C.~S.~Chen, C.~Q.~Geng, D.~Huang and L.~H.~Tsai,
  %``Multi high charged scalars in the LHC searches and Majorana neutrino mass generations,''
  Phys.\ Rev.\ D {\bf 87}, no. 7, 077702 (2013)
  [arXiv:1212.6208 [hep-ph]].
 


%%%%%%%%%%%%%%%%%%%%%%%%%%%%%%%%%%%%%%%%%%%%%%%%%%%%%
%%%%%%%%%%%%%%Type I 2HDM %%%%%%%%%%%%%%%
\bibitem{Barger:1989fj} 
  V.~D.~Barger, J.~L.~Hewett and R.~J.~N.~Phillips,
  %``New Constraints on the Charged Higgs Sector in Two Higgs Doublet Models,''
  Phys.\ Rev.\ D {\bf 41}, 3421 (1990).
  %%CITATION = PHRVA,D41,3421;%%
  
  

\bibitem{Xing:2002ta} 
  Z.~z.~Xing,
  %``Texture zeros and Majorana phases of the neutrino mass matrix,''
  Phys.\ Lett.\ B {\bf 530}, 159 (2002)
  [hep-ph/0201151].  
\bibitem{Xing:2002ap} 
  Z.~z.~Xing,
  %``A Full determination of the neutrino mass spectrum from two zero textures of the neutrino mass matrix,''
  Phys.\ Lett.\ B {\bf 539}, 85 (2002)
  [hep-ph/0205032].

%%%%%%%%%%Zero neutrino mass %%%%%%%%%%%%%%%%%%%%%%%%%%%  
\bibitem{Frampton:2002yf} 
  P.~H.~Frampton, S.~L.~Glashow and D.~Marfatia,
  %``Zeroes of the neutrino mass matrix,''
  Phys.\ Lett.\ B {\bf 536}, 79 (2002)
  [hep-ph/0201008]. 
\bibitem{Desai:2002sz} 
  B.~R.~Desai, D.~P.~Roy and A.~R.~Vaucher,
  %``Three neutrino mass matrices with two texture zeros,''
  Mod.\ Phys.\ Lett.\ A {\bf 18}, 1355 (2003)
  [hep-ph/0209035].
\bibitem{Guo:2002ei} 
  W.~l.~Guo and Z.~z.~Xing,
  %``Implications of the KamLAND measurement on the lepton flavor mixing matrix and the neutrino mass matrix,''
  Phys.\ Rev.\ D {\bf 67}, 053002 (2003)
  [hep-ph/0212142].
\bibitem{Honda:2003pg} 
  M.~Honda, S.~Kaneko and M.~Tanimoto,
  %``Prediction and its stability in neutrino mass matrix with two zeros,''
  JHEP {\bf 0309}, 028 (2003)
  [hep-ph/0303227].     
  
 \bibitem{Chau:1984fp} 
  L.~L.~Chau and W.~Y.~Keung,
  %``Comments on the Parametrization of the Kobayashi-Maskawa Matrix,''
  Phys.\ Rev.\ Lett.\  {\bf 53}, 1802 (1984).
  %%CITATION = PRLTA,53,1802;%%  
  


%%%%%%%%%%%%%%%%%%%%%%%%%%%LFV%%%%%%%%%%%%%%%%%%%%%%%%%%%%%%%%%

\bibitem{Adam:2013mnn} 
  J.~Adam {\it et al.}  [MEG Collaboration],
  %``New constraint on the existence of the $\mu^+ \to e^+\gamma$ decay,''
  Phys.\ Rev.\ Lett.\  {\bf 110}, 201801 (2013)
  [arXiv:1303.0754 [hep-ex]].

  
\bibitem{Nebot:2007bc} 
  M.~Nebot, J.~F.~Oliver, D.~Palao and A.~Santamaria,
  %``Prospects for the Zee-Babu Model at the CERN LHC and low energy experiments,''
  Phys.\ Rev.\ D {\bf 77}, 093013 (2008)
  [arXiv:0711.0483 [hep-ph]].  
%%%%%%%%%%%%%%%%%%%mu e conversion %%%%%%%%%%%%%%%%%%%%%%%%%%%
\bibitem{Bertl:2006up} 
  W.~H.~Bertl {\it et al.}  [SINDRUM II Collaboration],
  %``A Search for muon to electron conversion in muonic gold,''
  Eur.\ Phys.\ J.\ C {\bf 47}, 337 (2006).

\bibitem{Badertscher:1980bt} 
  A.~Badertscher, K.~Borer, G.~Czapek, A.~Fluckiger, H.~Hanni, B.~Hahn, E.~Hugentobler and H.~Kaspar {\it et al.},
  %``New Upper Limits for Muon - Electron Conversion in Sulfur,''
  Lett.\ Nuovo Cim.\  {\bf 28}, 401 (1980).
  %%CITATION = NCLTA,28,401;%%
  
\bibitem{Dohmen:1993mp} 
  C.~Dohmen {\it et al.}  [SINDRUM II. Collaboration],
  %``Test of lepton flavor conservation in mu ---> e conversion on titanium,''
  Phys.\ Lett.\ B {\bf 317}, 631 (1993).
  
\bibitem{Honecker:1996zf} 
  W.~Honecker {\it et al.}  [SINDRUM II Collaboration],
  %``Improved limit on the branching ratio of mu ---> e conversion on lead,''
  Phys.\ Rev.\ Lett.\  {\bf 76}, 200 (1996). 
  
\bibitem{Kitano:2002mt} 
  R.~Kitano, M.~Koike and Y.~Okada,
  %``Detailed calculation of lepton flavor violating muon electron conversion rate for various nuclei,''
  Phys.\ Rev.\ D {\bf 66}, 096002 (2002)
  [Erratum-ibid.\ D {\bf 76}, 059902 (2007)]
  [hep-ph/0203110].  
  
\bibitem{Ref:COMET}
Y.~Kuno et.al. [COMET collaboration], {\it An Experimental Search for lepton Flavor Violating $\mu - e$ Conversion at Sensitivity of $10^{-16}$ with a Slow-Extracted Bunched Beam}.

\bibitem{Ref:PRISM}
Y.~Kuno et.al. [PRISM/PRIME Group], Letter of Intent, {\it An Experimental Search for a $\mu - e$
Conversion at Sensitivity of the Order of $10^{-18}$ with a Highly Intense Muon Source: PRISM}.  
    %%%%%%%%%%%%%%%%%%%%%%%%%%%%g-2%%%%%%%%%%%%%%%%%%%%%%%%%


  
%%%%%%%%%%%%%%%%%%%%%%0\nu\beta\beta%%%%%%%%%%%%%%%%%%%%%%%%  
\bibitem{Pas:2000vn} 
  H.~Pas, M.~Hirsch, H.~V.~Klapdor-Kleingrothaus and S.~G.~Kovalenko,
  %``A Superformula for neutrinoless double beta decay. 2. The Short range part,''
  Phys.\ Lett.\ B {\bf 498}, 35 (2001)
  [hep-ph/0008182].
\bibitem{Deppisch:2012nb} 
  F.~F.~Deppisch, M.~Hirsch and H.~Pas,
  %``Neutrinoless Double Beta Decay and Physics Beyond the Standard Model,''
  J.\ Phys.\ G {\bf 39}, 124007 (2012)
  [arXiv:1208.0727 [hep-ph]].
  
\bibitem{Agostini:2013mzu}
  M.~Agostini {\it et al.}  [GERDA Collaboration],
  %``Results on Neutrinoless Double-$\beta$ Decay of $^{76}$Ge from Phase I of the GERDA Experiment,''
  Phys.\ Rev.\ Lett.\  {\bf 111}, no. 12, 122503 (2013)
  [arXiv:1307.4720 [nucl-ex]].
\bibitem{KamLANDZen}
  A.~Gando {\it et al.}  [KamLAND-Zen Collaboration],
  %``Measurement of the double-\beta decay half-life of ^{136}Xe with the KamLAND-Zen experiment,''
  Phys.\ Rev.\ C {\bf 85}, 045504 (2012)
  [arXiv:1201.4664 [hep-ex]].
\bibitem{Gando:2012zm}
  A.~Gando {\it et al.}  [KamLAND-Zen Collaboration],
  %``Limit on Neutrinoless $\beta\beta$ Decay of Xe-136 from the First Phase of KamLAND-Zen and Comparison with the Positive Claim in Ge-76,''
  Phys.\ Rev.\ Lett.\  {\bf 110}, no. 6, 062502 (2013)
  [arXiv:1211.3863 [hep-ex]].
\bibitem{Argyriades:2008pr}
  J.~Argyriades {\it et al.}  [NEMO Collaboration],
  %``Measurement of the Double Beta Decay Half-life of Nd-150 and Search for Neutrinoless Decay Modes with the NEMO-3 Detector,''
  Phys.\ Rev.\ C {\bf 80}, 032501 (2009)
  [arXiv:0810.0248 [hep-ex]].

\bibitem{Arnaboldi:2008ds}
  C.~Arnaboldi {\it et al.}  [CUORICINO Collaboration],
  %``Results from a search for the 0 neutrino beta beta-decay of Te-130,''
  Phys.\ Rev.\ C {\bf 78}, 035502 (2008)
  [arXiv:0802.3439 [hep-ex]].

\bibitem{Arnold:2005rz}
  R.~Arnold {\it et al.}  [NEMO Collaboration],
  %``First results of the search of neutrinoless double beta decay with the NEMO 3 detector,''
  Phys.\ Rev.\ Lett.\  {\bf 95}, 182302 (2005)
  [hep-ex/0507083].

\bibitem{Barabash:2010bd}
  A.~S.~Barabash {\it et al.}  [NEMO Collaboration],
  %``Investigation of double beta decay with the NEMO-3 detector,''
  Phys.\ Atom.\ Nucl.\  {\bf 74}, 312 (2011)
  [arXiv:1002.2862 [nucl-ex]].  
  




\bibitem{delAguila:2013mia} 
  F.~del Águila and M.~Chala,
  %``LHC bounds on Lepton Number Violation mediated by doubly and singly-charged scalars,''
  JHEP {\bf 1403}, 027 (2014)
  [arXiv:1311.1510 [hep-ph]].  
  
\bibitem{Aparici:2013xga} 
  A.~Aparici,
  %``Exotic properties of neutrinos using effective Lagrangians and specific models,''
  arXiv:1312.0554 [hep-ph].  
  
%%%%%%%%%%%%%%%%%%%%%%%%LHC%%%

\bibitem{atlas:2012gk}
  G.~Aad {\it et al.}  [ATLAS Collaboration],
  %``Observation of a new particle in the search for the Standard Model Higgs boson with the ATLAS detector at the LHC,''
  Phys.\ Lett.\ B {\bf 716}, 1 (2012)
  [arXiv:1207.7214 [hep-ex]].
  %%CITATION = ARXIV:1207.7214;%%

\bibitem{cms:2012gu}
  S.~Chatrchyan {\it et al.}  [CMS Collaboration],
  %``Observation of a new boson at a mass of 125 GeV with the CMS experiment at the LHC,''
  Phys.\ Lett.\ B {\bf 716}, 30 (2012)
  [arXiv:1207.7235 [hep-ex]].
  %%CITATION = ARXIV:1207.7235;%%  

%\bibitem{Lee:1972zzc} 
%  B.~W.~Lee,
%  %``Perspectives on theory of weak interactions,''
%  eConf C {\bf 720906V4}, 249 (1972).
  
%\bibitem{Barbieri:2006dq} 
%  R.~Barbieri, L.~J.~Hall and V.~S.~Rychkov,
%%  %``Improved naturalness with a heavy Higgs: An Alternative road to LHC physics,''
%  Phys.\ Rev.\ D {\bf 74}, 015007 (2006)
%  [hep-ph/0603188].



\end{thebibliography}
\end{document}